\documentclass{article}

\usepackage{PRIMEarxiv}
\usepackage{amsmath}
\usepackage[utf8]{inputenc} 
\usepackage[T1]{fontenc}    
\usepackage{hyperref}       
\usepackage{url}            
\usepackage{booktabs}       
\usepackage{amsfonts}       
\usepackage{nicefrac}       
\usepackage{microtype}      
\usepackage{lipsum}
\usepackage{fancyhdr}       
\usepackage{graphicx}       
\graphicspath{{media/}}     

\title{Empirical Determination of Baseball Eras: Multivariate Changepoint Analysis in Major League Baseball
}

\author{
  Mena CR Whalen, Gregory J Matthews \\
  Mathematics and Statistics\\
  Loyola University Chicago \\
  Chicago, IL\\
  \texttt{\{mwhalen3, gmatthews1\}@luc.edu} \\
   \And
  Brian M Mills \\
  Kinesiology and Health Education \\
  University of Texas at Austin \\
  Austin, TX\\
  \texttt{brian.mills@austin.utexas.edu} \\
}

\begin{document}
\maketitle

\begin{abstract}
We use multivariate change point analysis methods to identify not only mean shifts but also changes in variance across a wide array of statistical time series. Our primary objective is to empirically discern distinct eras in the evolution of baseball, shedding light on significant transformations in team performance and management strategies. We leverage a rich dataset comprising baseball statistics from the late 1800s to 2021, spanning over a century of the sport's history. Results confirm previous historical research, pinpointing well-known baseball eras, such as the Dead Ball Era, Integration Era, Steroid Era, and Post-Steroid Era. Moreover, the study delves into the detection of substantial changes in team performance, effectively identifying periods of both dynasties and collapses within a team's history. The multivariate change point analysis proves to be a valuable tool for understanding the intricate dynamics of baseball's evolution. The method offers a data-driven approach to unveil structural shifts in the sport's historical landscape, providing fresh insights into the impact of rule changes, player strategies, and external factors on baseball's evolution. This not only enhances our comprehension of baseball, showing more robust identification of eras than past univariate time series work, but also showcases the broader applicability of multivariate change point analysis in the domain of sports research and beyond.
\end{abstract}


\section{Introduction} 

The first professional baseball team in the United States, the Cincinnati Red Stockings, was formed in 1869 (\cite{BBHOF1869}). Many leagues came and went in the late 1800s, but the National League (NL), formed in 1876, emerged as the predominant league of the time. Near the turn of the century, the American League (AL) began growing in popularity and eventually reached an agreement with the NL to be the two major leagues of baseball with the winner of each league playing in the World Series starting in 1903.  

Throughout the long history of the game, baseball has gone through many changes and distinct eras, often classified qualitatively by historians.  \cite{Woltring2018} classify six eras of modern baseball: "Baseball has endured much change over the course of its history, and because of constant change, the modern era of baseball has been segmented into six distinct sub-eras. This common list is presented at Baseball-Reference, similarly depicting the eras as the Dead Ball Era (1901-1919), the Live Ball Era (1920-1941), the Integration Era (1942-1960), the Expansion Era (1961-1976), the Free Agency Era (1977-1993) and the Long Ball/Steroid Era (1994-2005)." For example, the time period between approximately 1900-1919, the "Dead Ball Era", was marked by low scoring games, few home runs, and dominant pitching. Alternatively, the "Steroid Era", lasting from approximately 1994 through 2005, was characterized by a rapid increase in power hitting often attributed to players using performance enhancing drugs. More recently adding to this list, Woltring et al. also identified and named a seventh era after 2006: the "Post Steroid Era." 

Given that the determination of eras tends to be inexact, a variety of approaches to identification of regimes between large changes can be helpful to understand innovations in gameplay, rules, league structure, team management, and athleticism. Historical, qualitative, and empirical quantitative approaches to this are complementary. Some academic work has therefore sought to \textit{empirically} identify structural changes in skills and gameplay, competitive balance, and fan demand (attendance). For example, evaluating univariate time series of performance measures, \cite{Groothius2017} "let the data speak", looking for structural change points over the period from 1871-2020. They analyzed four statistics: slugging percentage (SLG), home run (HR) rate, batting average (BA), and runs batted in (RBI) rate. For each of these statistics, they computed the mean and standard deviation across all players who had at least 100 at bats in a given season, yielding a univariate time series for each of these statistical measures. They then used the Lagrange Multiplier (LM) unit root test proposed in  \cite{LeeandStrazicich2003} to find change points, identifying two changes in slugging percentage in 1921 and 1992. The first change point marked the end of the Dead Ball Era, while the latter corresponds with the start of the Steroid Era. Similarly, \cite{Nieswiadomy2012}, tested for structural change points in a single player's performance: Barry Bonds. This work investigated a univariate time series of monthly On-Base-Plus-Slugging (OPS) of home run champion Barry Bonds over the course of his career, finding two large changes, one  in June of 1993 and another in September of 2000. The authors note that the latter occurred late in his career \--\ quite unexpectedly \--\ leading to speculation about steroid use by Bonds himself during the Steroid Era.

Additional work has evaluated the existence of structural changes in attendance and competitive balance. For example, \cite{LeeFort2005} and  \cite{FortLee2006} separately tested the stationarity of, and estimated structural changes in, competitive balance of the American and National Leagues from 1901 to 1999 using methods developed by \cite{baiperron1998, baiperron2003} and others (\cite{Andrews1993, bai1997, bai1999}). They measure competitive balance using the classical Noll-Scully ratio, adjusting for ties (\cite{Noll1988, Scully1989}). The analysis identified change points in competitive balance in 1912, 1926, and 1933 for the NL and in 1926 and 1957 in the AL, with improving trends in balance since this time. The authors largely attribute this to equalization of population centers (market sizes), television revenues, and increases in utilization of the international talent pool.

Baseball is not the only sport where this type of analysis has been applied. \cite{PalaciosHuerta2004} estimated structural change points in soccer using data from British soccer leagues through 1996. Notably, this work identified a change point in the mean of margin of victory in 1925 related to the change in the definition of offsides (changed from 3 players to 2 players). Additional change points occured in the variability of number of goals in the early 1980s and 1992, corresponding both with a change in number of points for a win (from 2 points to 3 points), and a change in the backpass rule. \cite{FortLee2007} looked for change points in competitive balance for three other major North American sports leagues: National Basketball Association (NBA), National Hockey League (NHL), and National Football League (NFL). They identified a number of change points in each sport that often, though not always, corresponded to league expansion, league mergers, or other major economic events in a sport (e.g. increased number of foreign players in the NBA in the late 1990s/early 2000s). 

Subsequent work has analyzed attendance shifts and trend changes in all four major North American leagues, seeking to understand relationships between league policies – such as free agency and expansion – on both balance and subsequent fan demand (\cite{LeeFort2008, MillsFort2014, MillsFort2018}). Many of the largest change points in attendance tended to be coincident with major wars, the Great Depression, league expansion, and labor disputes that resulted in cancelled games or seasons. Further empirical analysis has addressed competitive balance in college football and college basketball, which identified structural changes associated with realignment and a split into various divisions in National Collegiate Athletics Association (NCAA) (\cite{SalagaFort2017, MillsSalaga2015}).

All of this previous work focuses on change point analysis in a {\bf univariate} time series context. However, recent methodological developments in change point analysis allow for the estimation of change points in {\bf multivariate panel data}, which is the focus of the current manuscript. We use the Double CUSUM  (\cite{Cho2016}) and Sparsified Binary Segmentation algorithm (\cite{ChoFryzlewicz2014}) to identify change points in Major League Baseball (MLB) at multiple levels. We first seek to identify change points in a panel of league-level statistical measures and, separately, a team-level panel of individual statistical measures to empirically define different eras in baseball, providing a broader data-driven empirical context for historical analysis of the sport than past work. Subsequently, we search for change points within individual teams, using a panel of team-level statistics, to determine relevant eras of team performance and management. This latter analysis is used to empirically locate the initial emergence (and subsequent demise) of so-called "dynasties", periods of sustained excellent performance by a team.

\section{Methods}
\label{method}
Change point analysis is a statistical technique used to identify points where the statistical properties of a time series change. It is particularly useful in detecting structural breaks in time series data. In our study, we employ a variation on the cumulative sum (CUSUM) statistic, a well-established method for identifying such change points, as described by \cite{page1955test} and later extended by \cite{Cho2016}, \cite{ChoFryzlewicz2014}.

The CUSUM statistic is designed to detect shifts in the mean level of a time series. Imagine observing a series of data points over time; a change point occurs when the underlying process generating the data shifts, resulting in a noticeable difference before and after this point. For example, consider a baseball player's batting average throughout a season. If the player's average suddenly increases or decreases significantly at a certain point in the season, this would be a change point. The CUSUM method accumulates these potential structural changes over every time point, making it easier to identify significant deviations from the norm over the whole time series. This allows the data to demonstrate where these change events are occurring, outside of external input, leading to a more specific temporal location for underlying mean shifts. 

A change point can be found when the CUSUM, which can be thought of as a weighted difference between the two segments, is found to be significantly large compared to a threshold. The CUSUM statistic is calculated across the length of the time series and then broken down into smaller intervals and then calculating the CUSUM statistic for each of these intervals. These intervals are calculated cumulatively for every time point, say $b$, until a potential change point is found. If a change point is found this means that there is some significant breaking point, at a potential time point $b$, leading to a large difference before and after $b$. Then Binary Segmentation (BS) is used to break up the time series into segments based on the identification of the first change point (the largest CUSUM at a given $b$) in the length of the time series and then sequentially breaks down into smaller segments until no more change points are detected. The CUSUM statistic is found to be large enough for a change point to be determined based on a thresholding parameter related to the error term and the length of the time series (\cite{hinkley1971inference}). This would empirically detect a change in a player's batting average over the season versus a "hot-streak" of luck which is within the bounds of random chance.

This method is used in univariate analysis for the mean change of a time series but can have limitations in some situations. For instance, examining just one player's batting average does not account for anything outside of this player, like the overall team's performance or league performance. Our interest is in segmenting the structure of time series in multivariate settings for both mean and variance changes to understand the combined influence of change points across multivariate time series. Instead of every players' batting averages being analyzed separately for an individual players unique "hot-steak", an entire team would be analyzed together to determine how every players' batting average across a season determines if the entire team is experiencing an increased performance in batting. This is multivariate panel data, and investigating when changes occur across time series would require separate analyses for each one of the time series, resulting in multiple potential locations of change points with different thresholding parameters. Attempting to use a single threshold parameter can be done to remove some potential change points but can prove difficult at high dimensions and does not appropriately use information about \textit{shared change point locations}. Univariate analysis does not also allow for shifts in variance to be detected across multiple time series, potentially addressing if a team's batting average could be less variable over time centralizing around a certain level (or higher variability from widening batting averages across the team throughout the season). A univariate analysis of a single player's batting average can give no information about the variability of the team's batting average while multivariate variance change point analysis can highlight "stability" (or instability) within the team.

The methodology from \cite{Cho2016} proposed the Double CUSUM (DC) statistic which computes CUSUM statistics across the same intervals of all the time series and then creates a test statistic using \textit{ordered} CUSUM values within that interval. This is given in detail in the section 2 of \cite{Cho2016}. The DC enhances upon the univariate case by allowing for comparisons not only within a single time series but \textit{across} multiple time series. For example, instead of looking at just one player's batting average, we look at the batting averages of the entire team. The ordering of the CUSUM statistics prioritizes time series that have larger values since these would most likely have change points occurring within this interval and potentially in others. DC statistics across all intervals of each of the time series are then compared against a threshold to remove intervals that even at their largest value would not result in a change point. This thresholding process helps remove locations in time where there possibly is not a meaningful change point across \textit{all} of the time series.

We use the Sparsified Binary Segmentation (SBS) algorithm in combination with the DC statistics to identify change points across multiple time series by \cite{ChoFryzlewicz2014}. The SBS algorithm segments the time series panel after a change point, in mean or variance, is found to recursively find all change points in the panel. The sparsification comes from only considering certain DC values that are above a given threshold thus reducing insignificant contributions potentially reducing comparisons when dimensionality is high. The Generalized Dynamic Factor Model (GDFM) bootstrapping algorithm is employed to determine appropriate thresholds. For a detailed explanation, including the equations, the specifics of the bootstrapping algorithm, and how variance change points are calculated, please refer to \cite{Cho2016} section 4. Due to the nature of the methodology, evaluation of time series of differing lengths cannot be performed; all time series must be compared within the same period of time for calculating the CUSUM and comparing the same generic intervals.

We apply this methodology to multiple forms of panel data in baseball to evaluate and confirm changes in eras of the game for league-level statistics, within popular game metrics statistics, and determine dynasties within teams with team-level statistics.  R programming language (\cite{R2023language}) and the package "hdbinseg" (\cite{hdbinseg}) were used to wrangle, model, and visualize this analysis.

\subsection{Data}
We obtained seasonal level baseball statistics from the Lahman database from 1901 through 2021 \cite{Friendly2021Lahman} .We restrict our analysis to the period usually referred to as the "Modern Era" and focus on the American League (AL) and National League (NL) histories over this period (the AL began operation in 1901). Over the course of the sport's history, numerous statistics have been meticulously collected and maintained, which we leverage for our analysis, focusing on year-end statistics at the team- and league-level. Over the course of the sport's history, numerous statistics have been meticulously collected and maintained, which we leverage for our analysis, focusing on year-end statistics at the team- and league-level. We use this data to perform 3 different analyses using the same methodology to understand the different aspects and evolution of the sport. The first is understanding the overall eras of baseball using league-level data. These league-level statistics, includes all teams existing for each year in question from 1901 through 2021. The second form of analysis is about specific game measures evolving over time based on a panel of teams. For this team-panel analysis, it was necessary for each team to have existed from 1900 until the end of the data set in 2021, as the multivariate methodological approach requires equal length time series. This sample (a panel of team-level measures, estimated separately for each statistical measure) encompassed a total of 16 franchises, each presented in Table \ref{tab:franch}. The last form of analysis examines within a specific team to understand "dynasties" or "collapses" within a history of a franchise. For this within team-level analysis, a panel of statistical measures for each team is used, we include all existing teams across the franchise's entire existence, treating each as a single franchise irrespective of geographic moves or name changes. 

The statistics of interest included in the separate analysis are runs (R), hits (H), home runs (HR), base-on-balls (walks; BB), strikeouts (K), at-bats (AB), stolen bases (SB), number of games played in a season (G), runs against (RA), hits against (HA), home runs against (HRA), base-on-balls against (BBA), strikeouts against (SO), and home attendance totals (ATT). We use the G variable for total number of games for the season to transform all count statistics into per-game rate statistics. However, a few statistics, such as BB, SB, K, and ATT had missing data for certain years generally pre-1900. To address this, we imputed missing values with predicted values from a linear regression model using MICE (\cite{JSSv045i03}) due to low rates of missingness and to ensure consistent data across all analyses. The breakdown of these 3 analyses with descriptions of the variables included, panel dimension of the data, and years included can be seen in Table \ref{tab:analysis_overview}.

\begin{table}[]
\caption{Overview of the different analysis described in the paper organized by giving the section location, time period under consideration, the panel or multivariate dimension of the data, and number of analyses performed within each section of analysis.}
\begin{tabular}{|l|l|l|ll|l|}
\hline
Results Section & \begin{tabular}[c]{@{}l@{}}Number of \\ Analyses\end{tabular} & Analysis Description                                                                                       & \multicolumn{2}{l|}{\begin{tabular}[c]{@{}l@{}}Multivariate Dimension (Rows\\ of panel\end{tabular}}                                       & Years (Columns)                                                         \\ \hline
3.1     & 1                                                             & Overall Eras                                                                                   & \multicolumn{2}{l|}{\begin{tabular}[c]{@{}l@{}}Statistical Measures (4)\\ HR, SO, SB, BB\end{tabular}}                                  & 1900-2021                                                               \\ \hline
3.2     & 6                                                             & \begin{tabular}[c]{@{}l@{}}Eras of Baseball Metrics - \\ HR, SO, BB, \\ SB, R, and ATT\end{tabular} & \multicolumn{2}{l|}{\begin{tabular}[c]{@{}l@{}}Franchise (16)\\  Found in Table \ref{tab:franch}\end{tabular}}                                        & 1901-2021                                                               \\ \hline
3.3     & 30                                                            & \begin{tabular}[c]{@{}l@{}}Dynasties - Modern Franchise \\ (30)\end{tabular}                               & \multicolumn{2}{l|}{\begin{tabular}[c]{@{}l@{}}Statistical Measures (10)\\  R, H, HR, BB, K, RA,\\  HA, HRA, BBA, and SO\end{tabular}} & \begin{tabular}[c]{@{}l@{}}Beginning of \\ Franchise- 2021\end{tabular} \\ \hline
\end{tabular}
\label{tab:analysis_overview}
\end{table}

\subsection{Eras, Metrics, and Dynasties}
\subsubsection{Overall Eras and Era of Metrics}

Because we are interested in finding where in time different eras of baseball have emerged, we first focused the multivariate change point procedure to a league-level panel of time series on the 4 key seasonal league-average measures which have been analyzed in previous research: HR, K, BB, and SB.  These represent differences in approach with respect to power, contact, reaching base, and speed, respectively. Each measure was standardized by dividing the number of games in a season (G). The standardization was required due to the large changes in games played in MLB across this time period, ranging from 60 games in the 1800s to as many as 163 in recent years where a tiebreaker game was required at the end of the regular season. Here, the multivariate dimension in our analysis is the various statistical measures for each season (HR, SO, BB, and SB).

We next examined some key statistics individually, on how these statistics and their changes evolved, using the time series of all 16 teams. Those statistics included HR, SO, BB, SB, R, and ATT. These measures where divided by the number of games played in a season (G), to standardize differences across years similarly described in the the prior analysis. Unlike the prior league-level analysis, here the multivariate dimension of analysis is the 16 different time series of teams, with the change point procedure applied to each individual statistic. Teams included in this panel are included in Table \ref{tab:franch}.

\begin{table}
\centering
\caption{Franchise label and their modern team name for the 16 franchises that existed in every year in the interval 1901-2021, used for statistical measures analysis.}
\begin{tabular}{c|c}
\hline
Franchise Label & Current Franchise Name\\
\hline
ATL & Atlanta Braves\\
\hline
BAL & Baltimore Orioles\\
\hline
BOS & Boston Red Sox\\
\hline
CHC & Chicago Cubs\\
\hline
CHW & Chicago White Sox\\
\hline
CIN & Cincinnati Reds\\
\hline
CLE & Cleveland Guardians\\
\hline
DET & Detroit Tigers\\
\hline
LAD & Los Angeles Dodgers\\
\hline
MIN & Minnesota Twins\\
\hline
NYY & New York Yankees\\
\hline
OAK & Oakland Athletics\\
\hline
PHI & Philadelphia Phillies\\
\hline
PIT & Pittsburgh Pirates\\
\hline
SFG & San Francisco Giants\\
\hline
STL & St. Louis Cardinals\\
\hline
\end{tabular}
\label{tab:franch}
\end{table}

\subsubsection{Dynasties}

Continuing our investigation, we use change point analysis to empirically identify "dynasties" and "collapses" within modern baseball teams, irrespective of their length of a team's existence. For each team, we examined ten key statistics that represent a team's overall performance scoring and preventing runs (R, H, HR, BB, pitcher K, RA, HA, HRA, BBA, and batter SO) as a multivariate panel at the team-level. More specifically, the change point procedure was applied separately for each team panel, with 10 statistical time series for each team. Using this method, a dynasty would be identified when a team experiences a noticeable positive shift across their statistical outputs, whether hitting or pitching, whereas a collapse would be identified through negative shifts at a change point.

By scrutinizing these indicators, we aimed to uncover shifts in a team’s performance and assess their significance within the context of dynasties. These measures were standardized using the season average and standard deviation derived from all teams present in each respective season. By standardizing the measures, we capture a team’s offensive and defensive performance relative to peers within a season. Notably, if a team exhibits a high and positive offensive performance, it can be balanced to zero by large and negative defensive statistics when standardized. Therefore, if a team excels in both aspects of the game, it would have numerous time points above zero, whereas under-performing teams would display many below-zero values.

\section{Results}
\subsection{Eras: Overall Measures of Baseball}
We begin by describing change points identified using the league-level multivariate time series of standardized statistical measures. Figure 1 presents the estimated change points at the league-level, with 4 change points in mean, and 1 change point in variance of the multivariate time series of HR, SB, SO, and BB. The variance change point (1919) aligns closely with the earliest mean change point (1924), and appears to identify the end of the Dead Ball Era and beginning of the Live Ball Era. It is likely that variance change point began slightly earlier, as home run hitting in the Live Ball Era was largely kicked off by a few players or single player (Babe Ruth) in its early stages, leading to high variability in approaches. Subsequently, other players followed suit and swung for the fences after realizing the benefits to such an approach after recent changes to the way balls were used in games. As with past work, the end of the Dead Ball Era tends to be the most strongly empirically confirmed change point in MLB's history.

\begin{figure}
    \centering
    \includegraphics[width=\textwidth]{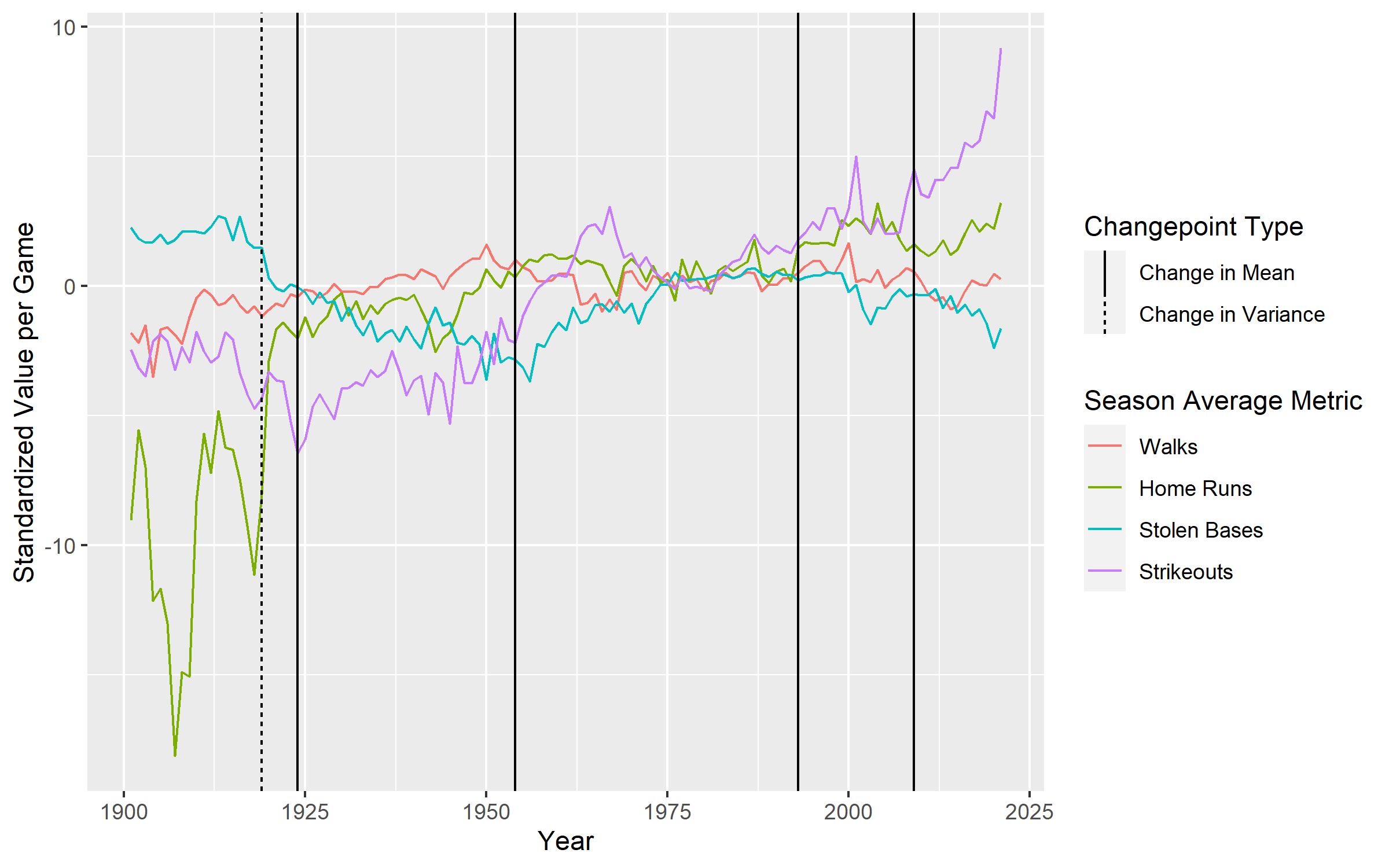}
    \label{fig:ave_ts}
    \caption{Results of overall eras analysis displaying standardized season average for the given statistics: Walks (BB), Home Runs (HR), Stolen Bases (SB), and Strikeouts (SO) along with their multivariate change points in mean and variance.}
\end{figure}

The next mean change occurred in 1954, concurrent with a sharp and continued increase in strikeouts. Stolen bases also began to increase after they had been decreasing since the start of the Deadball Era. \cite{mcmurray2015} describes the 1950s as "The Nadir of Stolen Bases." While not directly tied to traditionally separate eras, this tends to point toward a change in managerial philosophy, as well as possible umpire influences. Umpires during the 1950s were often accused of failing to enforce the set rule for pitchers, which made stealing bases particularly difficult. In 1950, there was also an increase in the size of the strike zone, and subsequent dramatic increases in strikeouts ensued throughout the decade. The combination of these differences seem to be the impetus for the detected change point around 1954, particularly as steals began increasing again during the 2nd half of the decade.

The combination of these differences seem to be the impetus for the detected change point around 1954, particularly as steals began increasing again during the 2nd half of the decade. The 1954 season occurred during the growth of integration, with a large influx of Black players to MLB after Jackie Robinson's 1947 debut. Many of these players were excellent base-stealers and were chosen by MLB teams in part for this skill. This may have brought novel play styles to the league. Indeed, among the top 25 career stolen base leaders in MLB, every post-integration player on the list is Black. While we hesitate to imply this is a race-specific skill, past work has shown that skill and task selection by teams themselves has had racial undertones. This seems to have been a driver of the SB changes that underlie the 1954 change point. \footnote{We thank a reviewer for pointing this out.}

The more recent changes, in 1993 and 2009, likely have straight forward interpretations. 1993 coincided with the early part of the Steroid Era. During this time, there was a rapid increase in home runs and strikeouts. From 1992 to 1993 alone there was a 23 percent increase in HR per game, marking an immediate jump over the course of one year. Strikeouts increased more gradually over this time and then accelerated near 2009, a time at which a long-increasing influx of Latin American players leveled off. This 1993 season also marked expansion to Miami and Colorado in the NL (and later to Tampa and Arizona), which potentially diluted the talent pool across the league. Although these teams do not appear in this league-level analysis, they could potentially influence output by other teams due to the low quality (including, again, a dramatic increase in strikeouts and walks). Further, the subsequent season marked one of the most consequential work stoppages in MLB history, canceling the 1994 World Series and some of the 1995 season. \footnote{The NL and AL were analyzed separately and the mean change points found in the overall eras analysis matched those change points found for each of the leagues.} 

Beginning around 2009, there was a well-documented expansion of the called strike zone by MLB umpires \--\ which continued through 2014 \--\ due to new monitoring and evaluation systems in place (\cite{mills2017a-labor}) that led to stark increases in strikeouts and reductions in walks and offense in general (\cite{mills2017b-econ}). As before, large changes to the relative rates of walks and strikeouts were again be a partial driver of this change point. In addition to the change to the strike zone, this period also identifies a time in which teams began increasing reliever usage, while decreasing innings pitched by both starting and relief pitchers. Although this was part of a longer decline that began before the turn of the century, the combination of these explanations seems likely to have driven the changepoint and its underlying drivers.

We exhibit the timing of estimated changes in Figure 1 alongside the individual time series of standardized key metrics (BB, HR, SB, and SO). Visually, many of the changes tend to be associated with batter strikeouts (SO), with large increases after the 1954, 1993, and 2009 mean change points, and to a lesser degree, home runs (HR). For the first change points in 1919 (variance) and 1924 (mean), the opposite was true: there was a sudden recent decrease in strikeouts a few years prior. Additionally, the first 3 mean change points coincide with a recent inversion of the relative rates of SB and HR, indicating particularly stark changes to game play during these eras. In all cases, these inversions were associated with strong increases in HR relative to SB.

\begin{figure}
    \centering
    \includegraphics[width=\textwidth]{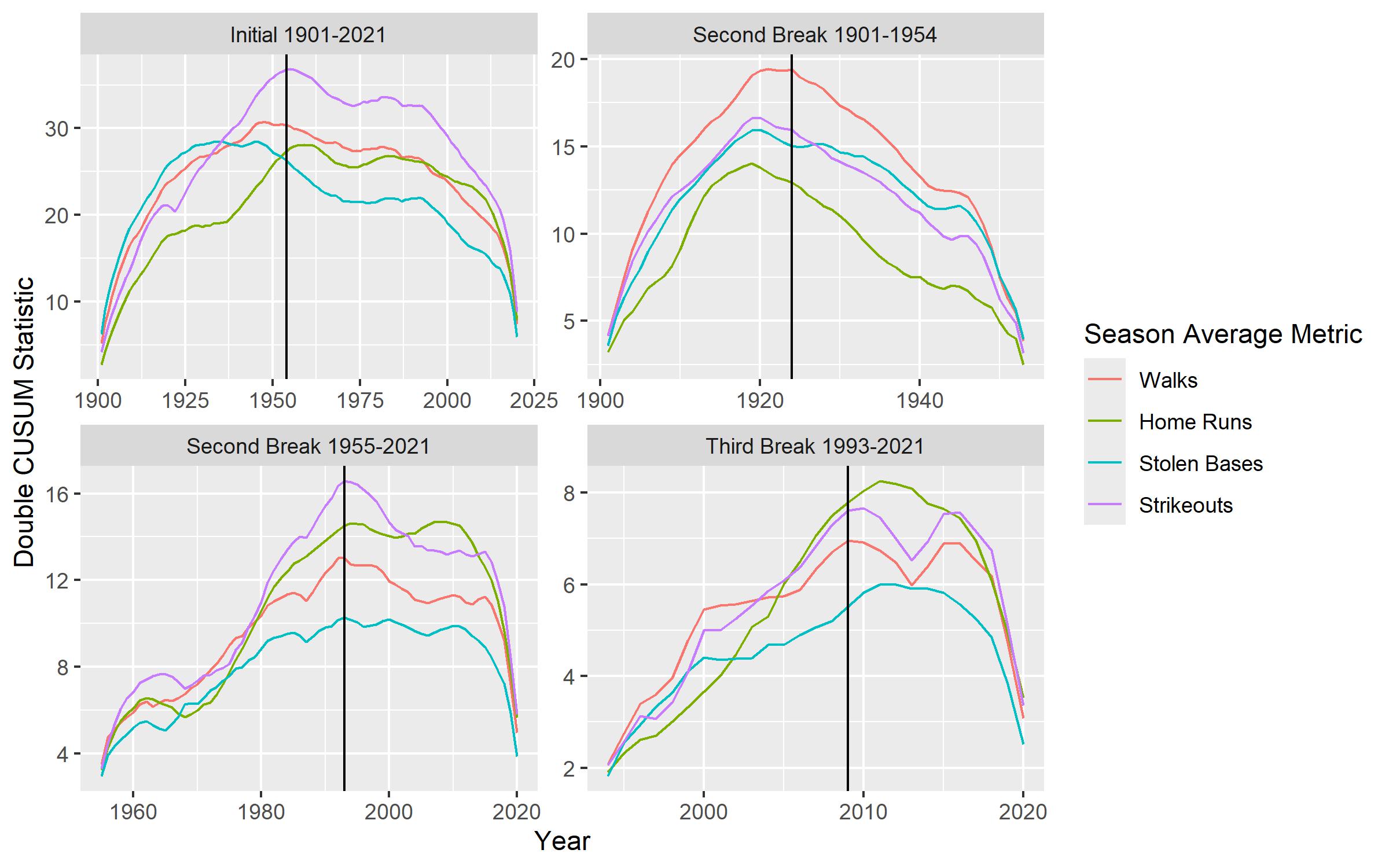}
    \label{fig:cusum_dc}=
    \caption{Double CUSUM values for each metric: Walks (BB), Home Runs (HR), Stolen Bases (SB), and Strikeouts (K) across each time period from the breaks in mean using the Sparsified Binary Segmentation Algorithm.}
\end{figure}

To fully investigate these changes in the mean, the DC statistic is examined for each metric to find which metric(s) are driving the potential change point locations. The first change point that is found in 1954, batter SO largely drives the change, while BB and and HR also have increasing DC statistics around this time. These peaks, seen in Figure 1, combine together to find the overall change point across all metrics to be at 1954 based on the largest values being greater than a given threshold. This highlights that multiple metrics are underlying a change point, as opposed to using univariate methods that would only express changes to one metric. 

The next step in the SBS algorithm breaks up the time series at the found change point to search for more change points from 1901-1954 and 1955-2021. In both sections of the break, a change point is found for 1924 and 1993, respectively. The earlier time period is mostly driven by BB with SO and SB following behind. These latter measures show peaks slightly earlier in time, but together highlight a change across all measures happening near the end of the Dead Ball Era. This peak in the DC statistic in BB comes from lower standardized BB before 1924 and afterwards BB is relatively constant around 0, showing very little changes in the metric after this point. The 1993 change point was driven by batter SO and HR with the other metrics similarly peaking around the same time period just at smaller values. Since this breaks is from 1955-2021, this is showing another increase in SO happening after 1993 that had been seen previously. The final change point found comes from the break between 1993-2021, estimating a change point around 2009, largely driven by HR and SO. These two metrics are increasing drastically after 2009 compared to before, and likely due to known changes in pitcher usage and batter focus on lifting the ball to hit more HR.

\subsection{Eras: Metric-Specific}

Here, we turn our attention now to the estimation of change points for individual statistics, leveraging the panel of team-level time series of each measure to assist in understanding drivers of the league-aggregate change points. We present mean and variance change points for each statistic in Table \ref{tab:mean:stat}. These are also presented visually in Figure 3.

\subsubsection{Attendance}

We begin with attendance to be able to compare directly to the abundant literature using univariate change point methods on attendance data in professional sports. Attendance and other statistical measures' change points can be found in Table \ref{tab:mean:stat} and Figure 3. We find similar change points to past work coinciding with the end of World War II (both for the mean in 1945, and the variance in 1944). Our analysis also identifies a change point in 1976, near the start of the Free Agency era and the expansion of the league to Toronto and (back to) Seattle. \cite{LeeFort2008} identify an attendance change point in the mid-1960s followed by a very strong upward trend - particularly in the American League - beginning shortly thereafter. While they did not estimate a structural change during the 1960s time for the National League, there is evidence that an increasing attendance trend seems to have begun near our estimated 1976 change point. Figure 3 also shows this steepening attendance trend around this time.  Finally, a variance change point was identified in 2008, which was the final year that baseball was played in both Yankee and Shea Stadium with both teams, the Yankees and the Mets, moving to stadiums with smaller capacities.   

\begin{table}[]
\centering
\caption{Change points present in statistics across the league, change point in mean unless noted as $^*$ for variance.}
\begin{tabular}{l|l}
\hline
Statistic & Changes\\
\hline
Attendance & 1944$^*$, 1945, 1976, 2008$^*$\\
\hline
Walks &   1936 \\
\hline
Home Runs & 1920, 1927$^*$,  1944$^*$, 1946, 1967, 1993,  1995$^*$, 2007$^*$\\
\hline
Runs & 1919, 1935$^*$, 1940 \\
\hline
Stolen Bases & 1919, 1919$^*$,  1966$^*$, 1967\\
\hline
Strikeouts & 1956, 1993, 1995$^*$, 2009 \\
\hline
\end{tabular}
\label{tab:mean:stat}
\end{table}

\subsubsection{Runs}
Our method identified two mean change points and one variance change point using the runs per game measure. The first mean change point in 1919 is a common finding in the literature and across some of our other statistics, coinciding with World War I, the Spanish Flu pandemic, and the end of the Dead Ball Era. The second mean change in 1940 is approximately concurrent with the end of the Live Ball Era and just prior to the start of the Integration Era. Lastly, the variance change point, which occurred in 1935, is near the peak of Live Ball Era scoring, suggesting that as run scoring came back down, variability in scoring was also reduced. In 1931, MLB instituted an official rule to standardize the height of the pitcher's mound at 15 inches. Mound height has been shown to have important impacts on run scoring, and this standardization may have been the reason for what appears to be reductions in the variability in run scoring across teams detected in 1935.

\subsubsection{Home Runs and Walks}
There are 4 estimated change points for HR for each of the mean and variance, which largely align with the traditionally noted eras. Mean change points approximately correspond with the start of the Live Ball Era (1920), the end of World War II and start Integration Era (1946), and the start of the Steroid Era (1993). The fourth mean change point was in 1967, just as the league lowered the pitcher's mound by 5 inches and just before shrinking the strike zone.  Variance change points are associated with the approximate start of the Integration Era (1944), the Steroid Era (1995), and the Post Steroid Era (2007). A fourth variance change point occurred in 1927 as the league shifted out of the Dead Ball Era and and into the Live Ball Era where a much larger emphasis was placed on power hitting. A mean change point in walks was identified in 1936. This change point occurs during the latter part of the Live Ball Era and is consistent with pitchers trying to avoid throwing pitches down the middle to home run hitters that were now more plentiful throughout the league. 

\begin{figure}
    \centering
    \includegraphics[width=\textwidth]{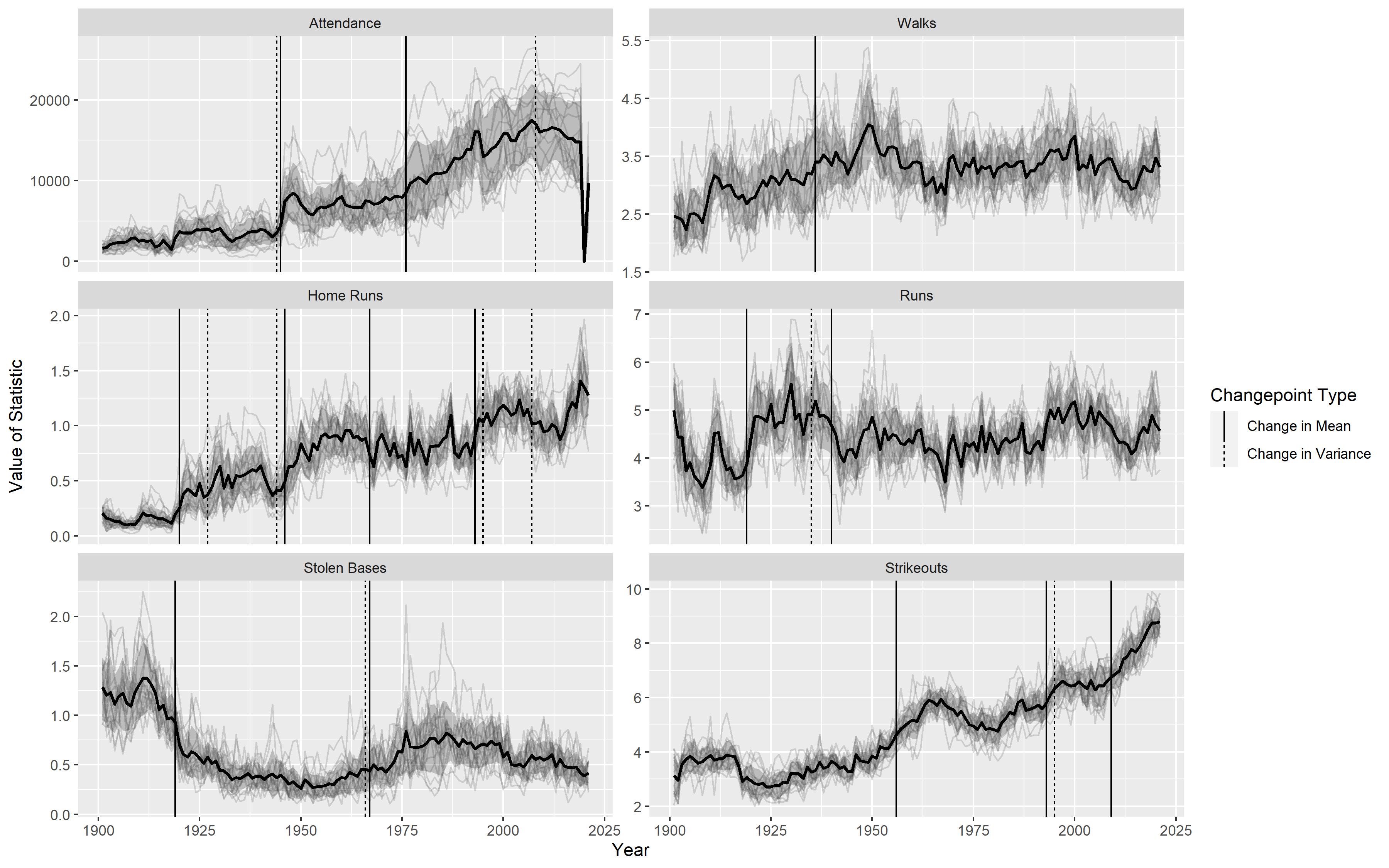}
    \label{fig:stats_ts}
        \caption{Change point of all metrics (Attendance, Walks, Home Runs, Runs, Stolen Bases, and Strikeouts) over the league for 120 years across 16 teams. The average value across all these teams is shown in bold with shaded standard deviation around it.}
\end{figure}

\subsubsection{Stolen Bases}
As offensive output trended toward hitting home runs in the Live Ball Era, the need for SB decreased (\cite{mcmurray2015b}).  This is seen by the estimation of a change point in the mean (associated with a drop) in SB in 1919.  In that same year, a change point in the variance was also observed as well.  In the latter half of the 1960s, another mean and variance change point pair is observed.  There is a variance change point in SB in 1966 and mean change point observed a year later in 1967. This was concurrent with a convergence in SB and HR rates that occurred beginning around this time and accelerating into the mid-1970s. \cite{mcmurray2015} notes that as larger parks began to open, Deadball Era play styles re-emerged in the early 1970s.

\subsubsection{Strikeouts}
The procedure detected 3 mean change points and 1 variance change point for strikeouts. Here, we note that two of the mean change points (and in combination with the variance change point) are associated with the start of traditional eras. The first mean change point in 1956 indicates a dramatic change in strikeouts (see Figure 3), which was later reduced by lower the pitching mound in 1968. The 1960s marked the Expansion Era, perhaps adding fuel to the strikeout increases during this time by increasing the number of players in the league (and subsequently diluting the talent pool in the short term). The 1993 mean change point (1995 variance change point) again approximately aligns with the start of the Steroid Era. The 2009 change point, while close to the start of the Post Steroid Era, also aligns with other work on changes to the umpire-called strike zone. Specifically,  \cite{mills2017a-labor},\cite{mills2017b-econ} shows that this expansion of the strike zone, which was concurrent with new umpire monitoring technology, accounted for as much as 40 percent of the changes to run scoring during this time through a decrease in the odds of contact on any pitch by 73 percent. This expansion began right after the 2009 season.

\subsection{Dynasties and Collapses: Modern Franchise}
Table \ref{tab:mean:team} exhibits the estimated change points for the team-level multivariate time series analyses used to identify team-specific eras across all statistics. These are also visualized in Figure 4. In this case, we identified change points for each team in the data using all team-specific statistical measures noted previously. Because there are a large number of change points across teams, we split our analysis by first consolidating common change point seasons across teams, and subsequently discuss particularly strong dynasties or collapses for relevant teams in the data. We leave inspection of remaining team-level change points to the reader.

In nearly every decade, there were changes points estimated that were common across teams, and consistent with some of the league-level changepoints discussed earlier. For example, 3 different teams (BOS, CIN, PHI) were estimated to have change points around 1918, near the end of the Dead Ball Era. Similarly there were 5 different teams (BOS, LAD, STL, CLE, DET) which experience change points shortly before or after the start of the Integration Era (these ranged from 1932 to 1944). There were also 4 teams (BAL, CLE, DET, MIN) with change points at the end of the Integration Era.  The most common single year associated with a change point was 1977, with 3 teams (LAA, MIL, SDP) experiencing change estimated changes at the start of the Free Agency Era.  The 1970-1974 time period.  The 1973-1978 time period also saw 4 other teams with estimated change points (CWS, SF, NYM, CIN). Additionally, 6 different teams were estimated to have change points near the start of the Steroid Era and around the 1994-1995 work stoppage (CLE, MIL, BAL, HOU, KCR, SEA). Finally, there were 4 teams (TB, BOS, HOU, WAS) with change points associated with the start of the Post Steroid Era, with change points ranging from 2007 to 2011.

\begin{table}
\centering
\caption{Change points for franchises in mean unless noted as $^*$ for variance.}
\begin{tabular}{l|l}
\hline
name & Changes\\
\hline
Arizona Diamondbacks & None\\
\hline
Atlanta Braves & 1887$^*$, 1990\\
\hline
Baltimore Orioles & 1959, 1999\\
\hline
Boston Red Sox & 1918, 1937,  2008$^*$\\
\hline
Chicago Cubs & 1891\\
\hline
Chicago White Sox & 1969$^*$, 1973\\
\hline
Cincinnati Reds & 1918, 1952, 1978 \\
\hline
Cleveland Guardians & 1944, 1958, 1966$^*$, 1993\\
\hline
Colorado Rockies & None\\
\hline
Detroit Tigers & 1912$^*$, 1942$^*$, 1958, 1988 \\
\hline
Houston Astros & 1999 ,  2010$*$\\
\hline
Kansas City Royals & 1997\\
\hline
Los Angeles Angels & 1977\\
\hline
Los Angeles Dodgers & 1940, 1961 \\
\hline
Miami Marlins & None\\
\hline
Milwaukee Brewers & 1977, 1996 \\
\hline
Minnesota Twins & 1959 \\
\hline
New York Mets & 1974$^*$\\
\hline
New York Yankees & 1918, 1964 \\
\hline
Oakland Athletics & None \\
\hline
Philadelphia Phillies & 1917, 1949 \\
\hline
Pittsburgh Pirates &  1902$^*$, 2001$^*$\\
\hline
San Diego Padres & 1977,  1977$^*$\\
\hline
San Francisco Giants & 1903, 1973\\
\hline
Seattle Mariners & 1999 \\
\hline
St. Louis Cardinals & 1932 \\
\hline
Tampa Bay Rays & 2007 \\
\hline
Texas Rangers & 1985 \\
\hline
Toronto Blue Jays & None\\
\hline
Washington Nationals & 2011\\
\hline
\end{tabular}
\label{tab:mean:team}
\end{table}

The change point detection was seemingly able to identify particularly strong shifts in performance of individual teams, and we focus on more recent change points here. In the 1990s, the Atlanta Braves (1990) and Cleveland Guardians (1993) quickly saw their fortunes change, culminating in multiple World Series berths for both teams and were seen as the two top teams dueling for the title of World Series Champion throughout the 1990s.\footnote{The similarity in their team mascots at the time likely heightened awareness of this cross-league rivalry.} While the Braves of this era are considered a dynasty by most baseball fans - headlined by their Hall of Fame pitching trifecta that included Greg Maddux, Tom Glavine, and John Smoltz - Cleveland is often remembered less fondly, given their inability to win a World Series during this stretch. Nevertheless, from 1994 to 2001, Cleveland was the best hitting team in baseball, leading the league in both on-base percentage and slugging percentage during this stretch. During this same peak of the Steroid Era, the Braves had a team earned run average of 3.55, leading the next closest team - the Los Angeles Dodgers (3.94) - by nearly half a run per game. This was arguably the most sustainably dominant pitching staff in the history of baseball. 

As both Atlanta and Cleveland saw the ends of their historic runs, the Seattle Mariners experienced a change point around the 1999 season. Seattle put together a team that had sustained success from 2000 through 2003, tying the all-time single season team wins record in 2001, which stands to this day. Most surprisingly, they managed this feat shortly after the departure of three of the greatest players in MLB history near the prime of their respective careers: Randy Johnson in 1998, Ken Griffey, Jr. in 2000, and Alex Rodriguez in 2001.

\begin{figure}
    \centering
    \includegraphics[width=\textwidth]{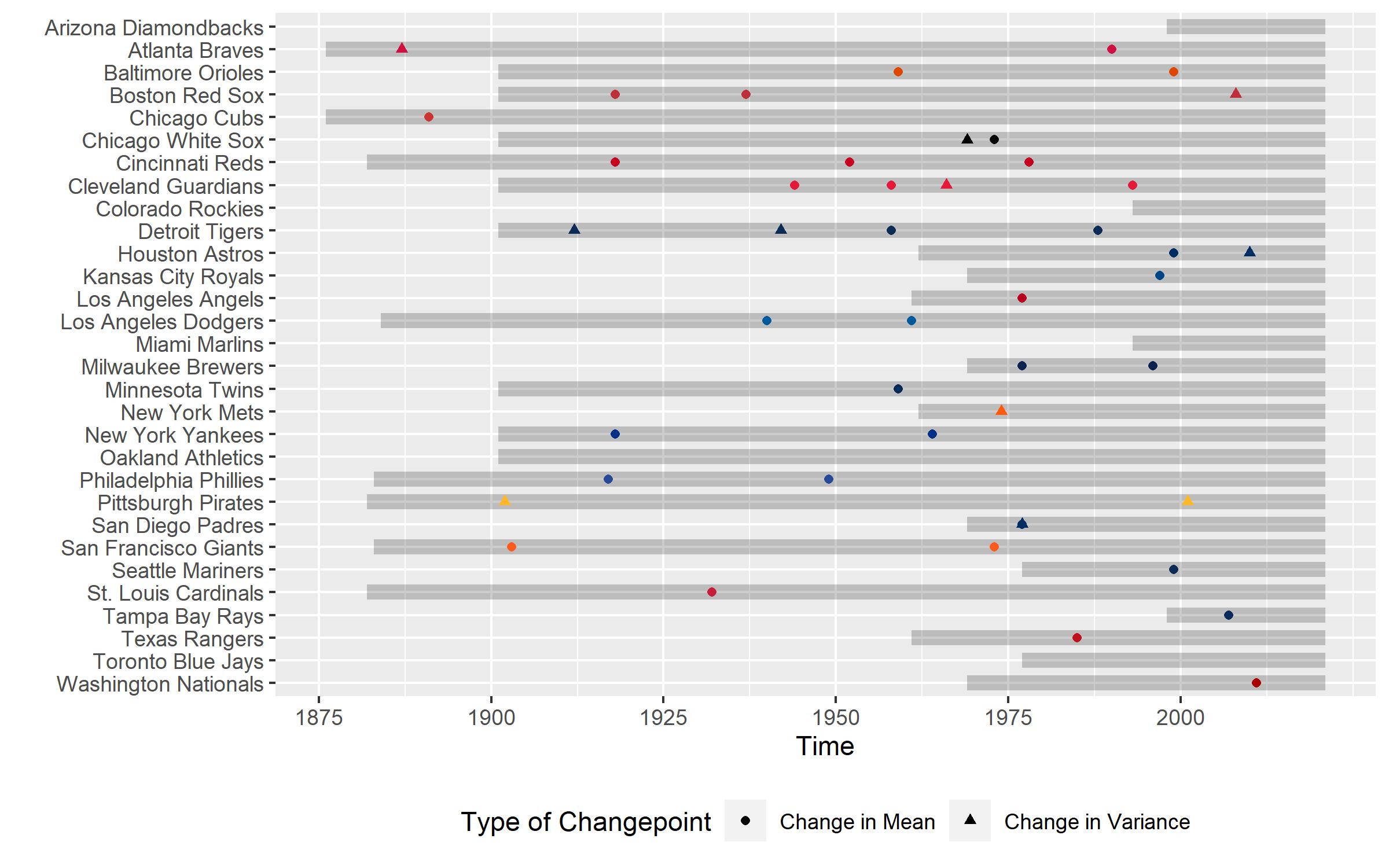}
    \label{fig:team_ts}

    \caption{Change point location in time for each franchise throughout their history, both in mean and variance if present. Gray bands represent the franchise's lifespan into the modern era.}
\end{figure}

More recently, the Tampa Bay Rays experienced a change point near the 2007 season, marking the last year of a streak of abysmal team records. From their inception in 1998 through the 2007 season, the Rays never won more than 70 games in a season, but suddenly vaulted to an American League East Division title - over the historically dynastic Yankees - and appeared in the World Series for the first time in franchise history. Since that season, the Rays have regularly been Division and Wild Card contenders, winning it 3 more times through 2021. Meanwhile, the Washington Nationals - which were moved from Montreal in 2005, shedding the Expos name - had a run of success starting in 2012 that culminated in a World Series title in 2019. This is likely identified by the 2011 change point in our data, and concurrent with the drafting of back-to-back first overall selections in the MLB draft (Stephen Strasburg and Bryce Harper). Prior to this, the Nationals had never had a winning season, and their predecessor in Montreal had only a single playoff appearance in their history. As a whole, the multivariate change point method seems to have identified these stark changes at the individual team level rather well.

\section{Conclusions}
Past work estimating structural changes to time series data in sports have exclusively focused upon univariate time series or separately estimating changes for different individual time series. We add to this literature by implementing advances to change point detection in the multivariate context, allowing for changes to take place not just in mean, but also in the variance across a panel of series. In our context, we were able to empirically identify traditional league eras directly from these multivariate data, including the Dead Ball Era, the Integration Era, the Steroid Era, and the Post-Steroid Era. This serves as strong support for the historical record as it relates to baseball eras. We also were able to classify when dynasties began, and identify declines for individual teams. These results are encouraging for the use of this method in sports data, and could be used to assist in historical description of leagues and teams across a number of dimensions that interest sports analysts and researchers.

Given the success of this approach here, we propose that other research questions related to sports leagues, teams, and management policies would be well served by the multivariate change point procedure. For example, there are opportunities to detect change points for individual athletes with multivariate time series. Past work has, again, focused on a single performance statistics at a time. However, it is now possible to identify structural changes to approach, power, speed, and game play for individual athletes across a variety of measures. Additionally, prior research has struggled to reconcile the diverse measurement of competitive balance in sports leagues, which is made up of various conceptualizations or components of balance and uncertainty (\cite{sanderson2002, MillsFort2014, gerrard2022}). Related work has focused on integrating various characteristics of balance into a single measure (\cite{humphreys2002}). Rather than separately focus on measures within their own series \--\ or aggregating into a single measure that loses information about certain dimensions \--\ this method could allow the change point detection in a collection of balance measure time series.

\newpage

\section{Supplemental Materials}

All code for reproducing the analyses in this paper is publicly available at github <provided after review>

\section*{Acknowledgement}
We thank Michael Lopez for suggesting we do "something with change point analysis".  

\section*{Funding}
  The authors received no financial support for the research, authorship, and/or publication of this article.
\bibliographystyle{unsrt}  
\bibliography{references}

\end{document}